\documentclass[a4paper,11pt]{article}
\usepackage[utf8]{inputenc}
\usepackage[english]{babel}
\usepackage{graphics}
\usepackage{graphicx} 
\usepackage{subfigure} 
\usepackage{multirow} 
\usepackage{geometry}
\usepackage{pdfpages}

\author{Andrea Celentano\thanks{andrea.celentano@ge.infn.it}\\ INFN - Genova }
\date{10 July 2014}

\usepackage[font=small,labelfont=bf]{caption}
\usepackage[style=numeric,sorting=none,backend=bibtex]{biblatex}
\bibliography{OptoTracker}

\newcommand{\projname}{OptoTracker }
\title{Research Project for INFN grant n. 16555:\\
\projname}

\begin{document}

\maketitle

\section{Project details}

\paragraph{Title of the project:} \projname

\paragraph{Name of the PI:} Andrea Celentano, PhD in Physics (March 2014)

\paragraph{Institution hosting the PI:} INFN, Sezione di Genova

\section{Short abstract}

The project \projname aims to investigate a new approach to track charged particles in a scintillating material, by using the optical signal. 
Our idea is to reconstruct the trajectory of a charged particle by collecting the scintillation light emitted along the path with pixelized photo-detectors. This would permit to obtain an image of the track, similarly to what is done in a photographic camera. Full 3D reconstruction is performed by using both the charge distribution and the hit time information folded in a sophisticated reconstruction algorithm.

This solution, compared to ``traditional'' tracking methods, exploits the fastest information carrier within a material: the light.
Therefore, an optical tracking detector would be intrinsically capable of sustaining a very high interaction rate. Moreover, the intrinsic resolution would not be limited by carriers diffusion, as happens in charge-transport based detectors.
This new technology could have a very large impact both on beam experiments, thanks to the possible increase in the acquisition rate, and in rare-physics experiments (double-$\beta$ decay, dark-matter search, neutrino oscillation search), where the enhanced particle-id and directionality capability can provide significant background reduction.

The project represents the first experimental investigation of this new technology. We plan to design and contruct an optical tracker prototype, and use it to characterize the different elements composing the detector: the scintillator, the photo-detector, and the reconstruction algorithm. Detailed MonteCarlo simulations will provide guidance in the prototype design, and will also be used to determine the components critical properties.
The final goal is to provide a proof-of-principle of this new technology, and to identify, for each component, which are the required characteristics and how these influence the detector performances.

\section{Scientific Background}

Tracking detectors play a major role in particle and nuclear physics experiments. 
Since the fabrication of the first cloud chamber in 1911 by Charles Wilson \cite{Wilson}, where charged particles were identified by looking at the different ionization tracks in the detector, huge improvements were made in the development of tracking detectors with higher read-out rate, higher resolution, and enhanced 3D reconstruction capabilities. 
Critical milestones in this process were: the gas-filled multi-wire ionization chamber by George Charpak in 1968 \cite{Charpak}, the drift chamber by Albert Walenta in 1971 \cite{Walenta}, and the gas-filled time projection chamber (TPC) by David Nygren in 1975 \cite{Nygren}. 


Other than in experiments at accelerators, tracking detectors are also widely used in rare searches experiments such as neutrino detection, double beta neutrinoless decay and dark matter (DM) search. In these experiments tracking detectors are used to measure and identify signal events trough their specific signature, rejecting the backgrounds. A significant number of direct DM search experiment employs noble gas TPCs in liquid phase. Measuring the direction of the scattered nucleus it is possible to reconstruct the direction of the incoming DM particle, to correlate it with the possible production sources and increase the background rejection capability. Two examples are the XENON100 experiment \cite{XENON100}, employing a 62 kg LXe TPC, and the DarkSide experiment \cite{DarkSide}, currently under construction, employing a LAr TPC with 50 kg active volume.

Particle tracking is even more critical in low-mass ($<$ GeV) dark-matter searches at accelerator beam-dumps, such as the recently proposed ``Beam Dump eXperiment'' (BDX) at Jefferson Laboratory \cite{BDX}. In these experiments, as opposed to direct-search experiments performed in mines or dedicated underground laboratories, the existing accelerator infrastructure constrains the detector location, and only limited shielding from cosmogenic particles can be constructed. Directionality can be exploited to identify events produced by the beam against the diffuse background expected from cosmogenic sources.  
 
Traditionally, charged particles tracking is performed by reading out a ionization signal. Secondary ionization electrons released by the primary particle along the path drift in an electric field reaching the collecting region, where they are detected by conductive wires (or other electrodes) producing a proportional signal. In micro-pattern gaseous detectors, electrons drift towards Micromegas structures \cite{micromega} or GEM-foils \cite{gem}. 
The transverse coordinates, perpendicular to the drift direction, can be measured by using at least two read-out planes or by segmentation of the collecting structure in pixels. 
The drift coordinate, instead, is reconstructed by measuring the drift time.

There are two main issues related to this approach. First, low cross-section experiments require detectors with large volumes and high density to reach a significant sensitivity to the signal. This results in non-trivial technical issues related to the operation of $\simeq m^3$ liquid volumes at cryogenic temperatures, maintaining sufficiently high levels of purity. Second, the position resolution is limited by the diffusion of charge carriers on their way through the detection area. These limitations make it difficult to obtain a sufficient position resolution in a detector with a large drift distance.
\\

\textbf{This project aims to investigate a new approach to measure trajectories of charged particles, by using the scintillation signal instead of the ionization signal.}
\\

When a charged particle moves across a scintillating material, optical photons are produced along the path. By measuring the emitted optical photons on a pixelized detector coupled to the scintillator, it is possible to reconstruct a bi-dimensional image of the particle track, as it is done in a photographic camera. However, instead of using optical devices (lenses and mirrors) to focus the image on the detector (``optical imaging''), we want to reconstruct tracks trough the measurement of the time of the hits on each pixel, with excellent resolution (``time imaging''). This would permit to reconstruct the full-track of the charged particle, performing a \textbf{truly optical 3D tracking.} 
Other than developing the required sensors and the associated read-out electronics, a critical part of this project is the development of the proper algorithms to perform optical tracking, i.e. to solve the time reversal problem associated with the measurement.  

\textbf{This new tracking approach is, in principle, applicable to all scintillating materials.} The project specifically investigates the case of organic scintillators.
This choice is motivated by the fact that organic scintillators have a fast decay time, a very high light yield, and a high light transmission. As discussed later, these properties make the time reversal problem easier to solve. 

\section{State of the art}

Organic scintillator-based detectors, both plastics and liquids, are today very common in physics experiments, thanks to the ease of use, the reduced cost, and the possibility to be manufactured in different sizes and shapes. The ``traditional'' setup of a plastic scintillator counter includes a photo-detector on one side, to collect the scintillation light and produce an electric pulse whose amplitude is proportional to the energy deposited in the detector by ionizing radiation. 

Recently, it was shown that it is possible to reconstruct the trajectory of a charged particle within a plastic scintillator, by measuring the scintillation signal with a highly segmented photo-detector \cite{scint1}.
In this work, the authors presented a ``proof-of-principle'' of the optical tracking approach, by using a small (4 mm $\times$ 4 mm $\times$ 4 mm) plastic scintillator cube coupled to a Hybrid PMT \cite{HPMT}. The image focusing was obtained by means of optical lenses and mirrors. The achieved position resolution was $\simeq 30 \, \mu$m. This pioneering study represents the state-of-the-art result in this field, and demonstrates the high potential of the new approach. The performances of the system were, however, limited by the use of optical devices for image manipulation and focusing: sharp tracks were observed only from an inner part of the scintillator, close to the focal plane. 

We want to further investigate the tracking capabilities of scintillating materials, exploiting also the time information and, therefore, avoiding the use of optical mirrors and lenses to ``focus'' the image on the detector. We expect that, with this new approach, we can achieve comparable performances, and overcome the limitations imposed by the optical focusing. 
\textbf{The design of an optical tracking  detector requires state-of-the-art technologies, with extreme performance in all of the involved components.}

\begin{itemize}
\item{A fast, high light-yield, highly transparent scintillator is required. In particular, the fast decay time is a mandatory characteristic for this application. In fact, photons in the scintillator are produced in a time-stochastic process, related to the fluorescent molecules de-excitation, with exponential decay time $\tau$. This induces a dilution in the correlation between the photon hit time measured by the detector and the corresponding production time in the scintillator. 

The light-yield also plays a role in the process: the higher it is, the higher the probability that the first photon hitting a pixel in the photo-detector belongs to the edge of the exponential decay distribution. 

Finally, the optical properties of the scintillator are critical: a very high transparency and a high optical uniformity are required, to minimize internal diffusion and to guarantee that the image on the photo-detector is sharp and not distorted.

A very large number of organic scintillators, both plastics and liquids, are today available, with different characteristics. Furthermore, known solutions exist to modify the properties of these materials, by altering their chemical structure. For example, to decrease the decay time, it is possible to dope the scintillator with a quencher: the use of benzophenone or acetophenone, in different concentrations, can lower $\tau$ down to $\simeq 300$ ps, at the price of reduced light-yield\cite{doping}.

}

\item{The main requirements for the photo-detector to be coupled to the scintillator are the sensitivity to single photo-electrons, an excellent timing resolution (better than 100 ps for a single photo-electron), a high segmentation, and finally a spectral sensitivity compatible with the scintillator emission range. 

The state-of-the-art detectors with these characteristics are the ``Large Area Picosecond Photo-Detectors'' (LA-PPDs), currently under development. This is a big effort to design a new state-of-the-art photo-detector with extreme time resolution (better than 30 ps per single photo-electron), high sensitive area (a $20\times 20$~cm$^2$ prototype has been already manufactured), and very fine pixelization (20 $\mu m$) \cite{lappd}. Having successfully completed in 2013 an initial 3 year pilot project, where many of the underlying technologies have been placed on a solid initial footing, the upcoming years will further develop and fully integrate these elements, culminating in the operation of first test detectors. 


}

\item{A fast, low-noise, multi-channel readout system is required. The system is used to elaborate the analog signals from the photo-detectors and extract from them the quantities of interest, i.e. the number of photons per pixel and the associated time. 
Given the very high number of channels to be acquired, the readout system must be developed using a compact Application-Specific Integrated Circuit (ASIC). The main requirements are an optimal intrinsic time resolution, better than the detector one, and a low intrinsic noise, to measure single photo-electron signals. Other requirements are the possibility to fine tune each sensor channel gain independently, and to be operable in self-trigger mode. 
}

\item{Finally, a robust and mature time reversal imaging algorithm is needed. The algorithm should unambiguously identify the track of a charged particle in the scintillator from the position and time of the hits in the photo-detectors or, in case of a point-like interaction (such as a low energy recoil proton from a neutron scattering), identify the interaction point. 

}
\end{itemize}

\section{Objectives of the proposal}

The project aims to use the scintillation signal to reconstruct the trajectory of ionizing particles within a scintillator, by detecting the emitted optical photons with highly segmented and fast photo-detectors. Three-dimensional track reconstruction is performed by measuring the time of the photon hits on each pixel, and then applying an advanced time reversal imaging algorithm.

\textbf{Preliminary MonteCarlo simulations showed that the new tracking approach is very promising.} In the simulations, the tracking detector was modeled as a $5 \times 5 \times 5$ cm$^3$ plastic scintillator cube, with light yield $10^4 \: \gamma/MeV$ and decay time $\tau=1.4$ ns, coupled to a photo-detector on each face. Each photo-detector was a $16 \times 16$ matrix of pixels, with $25\%$ quantum efficiency and $50$ ps single photo-electron time resolution. 

\begin{figure}[tpb]
\centering
\subfigure{\includegraphics[width=0.44\textwidth]{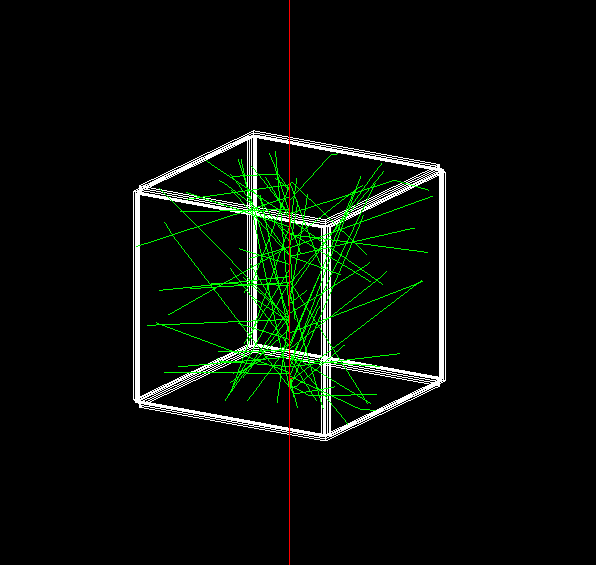}}\quad
\subfigure{\includegraphics[width=0.44\textwidth]{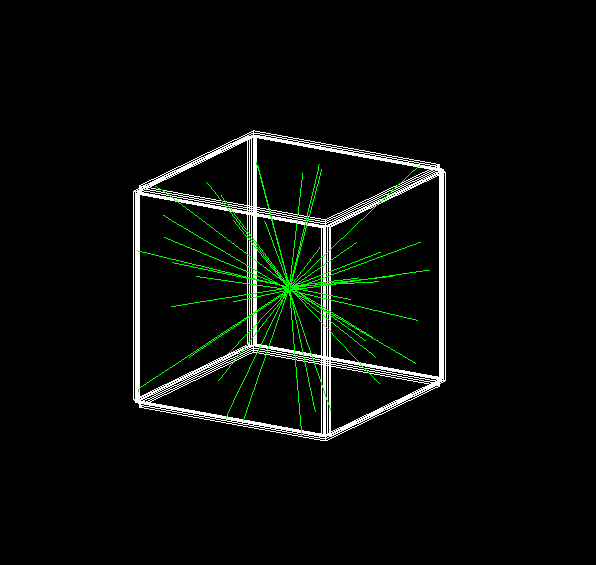}}
\caption{\small \label{fig:confronto} Comparison of the scintillation light emitted by a 2 GeV $\mu^{-}$ crossing the scintillator volume (left) and by a 10 MeV proton from a neutron scattering in the scintillator center (right).}
\end{figure}

\begin{figure}[h!tpb]
\centering
\subfigure{\includegraphics[width=0.48\textwidth]{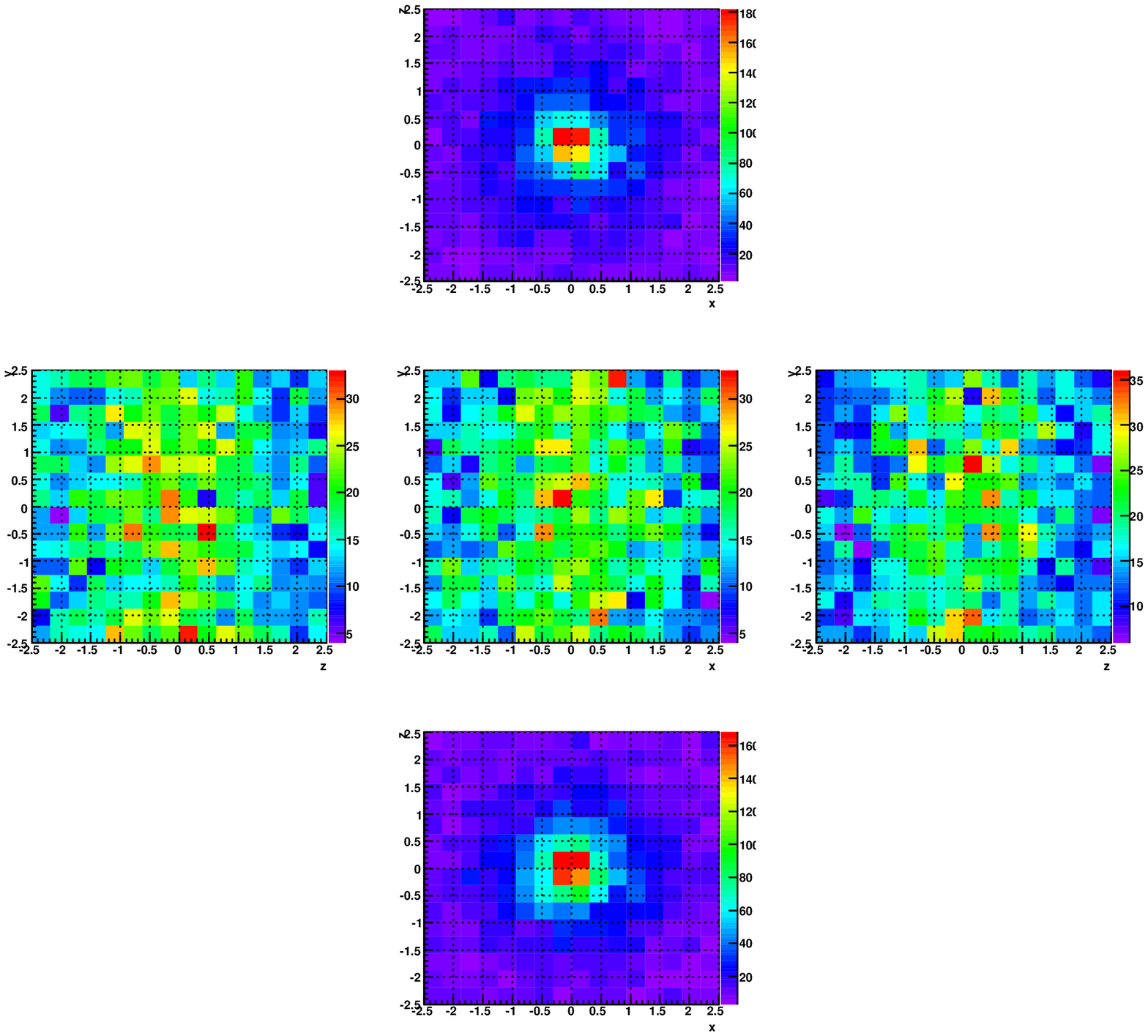}}\quad
\subfigure{\includegraphics[width=0.48\textwidth]{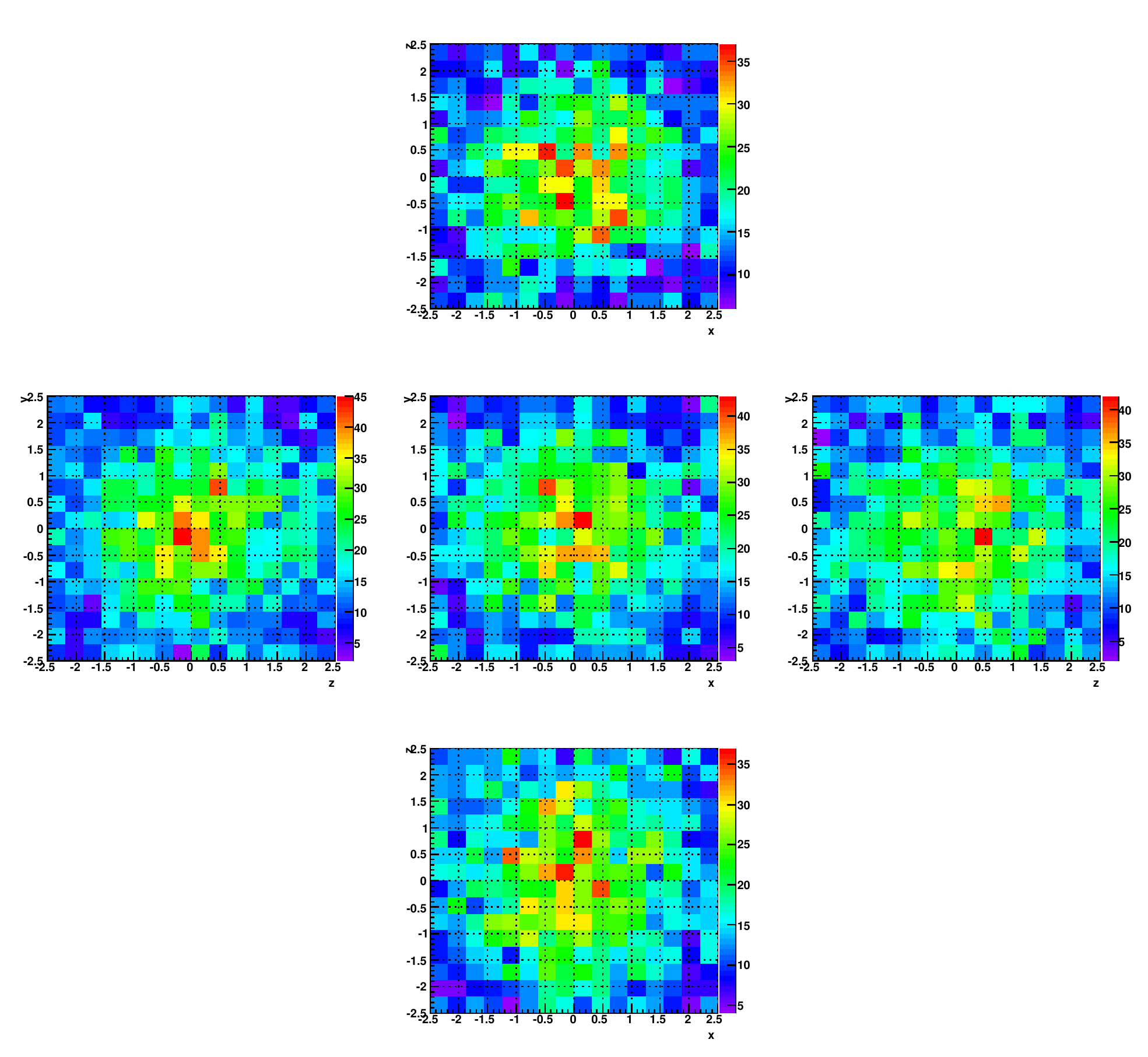}}
\caption{\small \label{fig:confronto1} Comparison of the detector response to the scintillation light emitted by a 2 GeV $\mu^{-}$ crossing the scintillator volume (left) and by a 10 MeV proton from a neutron scattering in the scintillator center (right). Each histogram corresponds to a detector face, and reports the response of the photo-detector there coupled in terms of the hit numbers distribution for each photo-detector pixel. For graphical reasons, only 3 of the 4 lateral faces are reported.}
\end{figure}

The response of the detector to different particles was simulated. In particular, we simulated muons passing trough the detector and neutrons hitting a proton that then recoiled, releasing a visible amount of energy. For each event, the number of hits on each pixel, as well as the first hit time, were recorded. A qualitative but very clear example is shown in Figure~\ref{fig:confronto}, comparing the optical emission of a crossing muon (left) and a recoil proton in the detector center (right). The corresponding hit number distribution is shown in Figure~\ref{fig:confronto1}. These two cases produce a very different hit pattern, that can be clearly distinguished.

Other than showing that it is possible to distinguish crossing particles from short-range particles (such as protons scattered by a neutron), the preliminary simulations indicated that two different muon tracks, with impact points at $\simeq 2$~mm distance, can be resolved from the hit pattern, as well as protons scattered in the detector within $\simeq 2$~mm distance. Finally, the hits time distribution permits to distinguish the direction of a crossing particle, ``downgoing'' or ``upgoing''. 

\textbf{This is only a very crude estimate of the resolution that can be achieved with this new technology applied to a large-scale detector.}
\\

\section{Description of the proposed research activity}

\textbf{During the project we will build an optical tracker prototype to study how the different components affect the ultimate performances of the whole system}.
We will measure the prototype response using variuos configurations, i.e. employing different organic scintillators and photo-detectors.
These measurements will be supported by detailed MonteCarlo simulations, to identify the best prototype configuration (for example, for a given dimension of the scintillator, the best pixelization of the photo-detector), and finally to determine, for a given setup, which are the expected performances. 


The project will be structured in Work Packages, each with a definite task, goal, and time-schedule. 
The proposed research activity will be mainly performed in the laboratories of the ``Gruppo 3'' research group of INFN Genova. This will permit to use the research infrastructures available there and to collaborate with the technical departments (see also the attached endorsement letter from Professor Sandro Squarcia, INFN-Genova director, who supports the use of the INFN-Genova resources and instrumentation).

\newpage
\subsection{Working packages description}

\subsubsection{WP1: development of MonteCarlo simulations} 
The first Working Package will be devoted to the development of a detailed MonteCarlo simulation of a possible optical tracking detector setup, with an organic scintillator coupled to pixelized photo-detectors, using the Geant4 framework. The simulation will be used to determine the performances of a given detector configuration, and, therefore, to define the optimal characteristics of the involved components, such as the geometry, the scintillator properties, the detector pixelization and resolution.   

We will start by investigating how optical photons are simulated within Geant4, focusing, in particular, on how the detector and the optical surfaces inhomogeneities are described. These effects are critical in this application, where a correct description of the light propagation is essential. 

MonteCarlo simulations will be validated by comparing the corresponding results with the experimental data measured with the detector prototype. The goal is to define a set of clear observables that can be measured, such as the hit distribution on the photo-detector pixels after the passage of a cosmic muon, and compare these with the simulation. 

Eventually, we will use the data measured with the prototype to optimize the MonteCarlo simulations, introducing in Geant4 a more complete description of optical processes, with free parameters to be tuned from experimental results.

The main tasks of this working package are:
\begin{description}
\item[WP1.1: investigation of optical transport in Geant4 (months 1-2).]{The first months will be employed for advanced training on Geant4, learning how optical processes are implemented. In particular, we will focus on how possible inhomogeneities and defects are described.}
\item[WP1.2: development of MonteCarlo simulations (months 3-6).]{}
\item[WP1.3: validation of MonteCarlo simulations (months 12-16).]{}
\item[WP1.4: optimization of MonteCarlo simulations (months 17-24).]{}
\end{description}

The milestones of this working package are:
\begin{description}
\item[WP1.1 (month 6):]{MonteCarlo framework for detector simulation developed.}
\end{description}

\newpage
\subsubsection{WP2: study and development of the optical tracking algorithm}

In this Working Package we will investigate the algorithms required to perform the optical tracking, i.e. to solve the time-reversal problem associated with this application.

First, we will define the problem, identifying which information can be obtained by the measurement, which are the properties of the detector that must be known (such as the geometry, the optical characteristics, \ldots), and understanding which are the possible ambiguities.
Then, we will develop the algorithm by investigating inversion methods employed in other fields, studying if they are adaptable to this specific problem. 

In particular, our project shares common issues with the field of Optical Tomograpy: the definition of the system response function, the backprojection algorithms, and the noise effects. Therefore, we will focus on the methods employed there, based on the Expectation-Maximization approach \cite{EM}. We will try to adapt them to our new approach, that also exploits the hit-time information in the reconstruction algorithm.  
The different algorithms will be tested and optimized using MonteCarlo simulations, and then validated with data collected with the prototype.

The main tasks of this working package are:
\begin{description}
\item[WP2.1: definition of the time-reversal problem (months 1-2).]{The first task of the working packages is to define the time-reversal problem, in terms of the input data available from the measurement, the required solution, and the possible associated ambiguities.}
\item[WP2.2: investigation of time-reversal algorithms in other fields (months 3-4).]{Before starting the algorithm development, we will study how analogous problems are handled in other fields, to see if we can adapt the corresponding algorithms to our needs.}
\item[WP2.3: development of the first version of the optical tracking algorithm (months 5-9).]{}
\item[WP2.4: test of the optical tracking algorithm (months 10-18).]{}
\item[WP2.5: optimization of the optical tracking algorithm (months 19-24).]{}
\end{description}

The milestones of this working package are:
\begin{description}
\item[WP2.1 (month 9):]{first version of the optical tracking algorithm developed.}
\end{description}

\newpage
\subsubsection{WP3: experimental investigation of optical tracking.}

In this Working Package we will investigate the different components that are needed for an optical tracking detector: the scintillator, the photo-sensors, the tracking algorithm. We want to understand how the behavior and the performances of the system are affected by the components properties, and systematically compare the results with those obtained from the MonteCarlo simulations. A non-exhaustive list of points that we want to investigate includes: the detector geometry (size and shape of the scintillator), the scintillator material, the photo-detector pixelization and resolution, and the reconstruction algorithm strategy. 

\textbf{To perform these studies, we will design and build a prototype of an optical tracking detector.} This will be the first example of a detector optimized for the new tracking approach and, therefore, it will permit to study and solve all the issues related to the new technology, that are not fully described by the MonteCarlo simulation, such as: the detector inhomogeneities, the optical coupling between the scintillator and the photo-sensor, the electronic noise. 

We will start the prototype design from a configuration suggested by MonteCarlo simulations. The prototype will be used, in turn, to validate the simulations and to perform a first test of the optical tracking. 
Good candidates for the plastic scintillator and the photo-multipliers are, respectively, the BC-420 from Saint Gobain, with $1.4$ ns decay time, light yield $\simeq 1.3\cdot 10^4 \: \gamma/\mbox{MeV}$, and attenuation length greater than $1$ m, and the Hamamatsu H9500 model, with 256 pixels (16x16) on a sensitive area of $\simeq 5$~cm$^2$. 

We will then modify the prototype configuration, using different components, to measure how these reflects in the detector behavior. We plan to change the scintillator geometry (size and shape), and to test different materials, both plastics and liquids. If necessary, we may also consider to produce a custom organic scintillator with improved properties, specifically for this project. A research unit with this specific task is present in the project.
For each scintillator configuration, we will try different photo-detectors, exploring, for example, the use of multi-anode photo-multipliers and silicon photo-multipliers (SiPMs), with different size, pixelization and resolution. SiPMs are a very promising solution: they have an enhanced quantum efficiency and single photo-electron detection efficiency compared to traditional photo-multipliers, and can easily reach 50 ps time resolution for single photons detection \cite{sipm}. During these studies, MonteCarlo simulations will be used as a guidance to determine, for each scintillator configuration, the most suited photo-detector choice to start with, and to determine what the foreseen performances are.

Concerning the electronic readout system, we plan to use existing state-of-the-art ASICs for multiple channels readout, originally developed for other applications. Two promising candidates are the MAROC3 ASIC developed at the IN2P3 ``$\Omega$meg$\alpha$'' laboratory  \cite{maroc3}, that represents a very a good starting candidate, and the new TOFPET ASIC, originally developed for Time-of-Flight Positron Emission Tomography (TOF PET) imaging \cite{tofpet}, with higher timing performances. Both ASICs are compatible with SiPMs and MA-PMTs. The readout system will be designed and built in collaboration with the INFN-Genova electronics department, whose Director (Ing. Paolo Musico) already expressed his interest and willingness to support the project.

The main tasks of this working package are:
\begin{description}
\item[WP3.1: definition and design of the first prototype configuration (months 5-8).]{The initial task of the Working Package will be to determine the first configuration of the optical tracker prototype, trough MonteCarlo simulations, and then to proceed in the design.}
\item[WP3.2: design and construction of the readout electronics system (months 7-11).]{The readout system will be designed to match the prototype configuration identified in the previous task. We will realize a versatile system, that can be used also with the other prototype configurations that we will build and characterize during the project.}
\item[WP3.3: construction of the first optical tracker prototype (months 9-12).]{}
\item[WP3.4: test of the first  optical tracker prototype (months 13-15).]{The response of the prototype to different particles will be measured. The first tests will be performed at INFN-Genova with cosmic muons. The possibility to perform other tests in dedicated facilities, such as the INFN-Frascati BTF for electron beams will also be considered.}
\item[WP3.5: data analysis and comparison with MonteCarlo results (months 16-18).]{The data measured during the prototype tests will be extensively analyzed, employing the optical tracking algorithm, \textbf{with the goal of providing the first, experimental proof-of-principle of this new technology.} Data will also be used to validate MonteCarlo simulations.} 
\item[WP3.6: experimental characterization of the optical tracker components (months 19-24).]{\textbf{This is the most critical and challenging task of this Working Package.} We will change the prototype configuration to measure how the different components influence the behavior and the performances. Given the exploratory nature of this task, we decided to maintain it sufficiently general, without yet a defined order of configurations that will be explored. This will be defined after the experience gained with the first prototype tests, using MonteCarlo simulations a guidance.}
\end{description}

The milestones of this working package are:
\begin{description}
\item[WP3.1 (month 8):]{first prototype version defined and design completed.}
\item[WP3.2 (month 12):]{first prototype built and ready for tests.}
\item[WP3.3 (month 18):]{data from the first prototype analyzed.}
\end{description}

\newpage

\subsection{Composition of the research team}

The main research unit of this project will operate at INFN-Genova, under the PI coordination. This will permit to use the research facilities there installed, and to collaborate with the technical departments on specific tasks: the electronics department for the design of the readout system and the mechanical workshop for the design and construction of the prototype structure. Since the PI is well integrated within the INFN-Genova ``Gruppo 3'' research group, where he has performed his previous research activity (PhD, post-doc), the project will strongly benefit of the expertise of this team.

Given the exploratory nature of this project, involving also a possible study of new materials with specific properties, a second research unit from the Chemistry department of Genoa University will be part of the project.

\subsubsection{INFN-Ge unit}

This unit operates under the coordination of the PI, who will take care of the overall scientific strategy and organization of the project. The INFN-Ge unit will have a major role in each Working Package. This unit is in charge of developing the MonteCarlo simulations, studying and implement a first version of the optical tracking algorithm, and building the detector prototype to perform the measurement campaign on the different involved components.  

During the first months of the project, the PI will possibly hire a PhD student to be involved in the project.

\paragraph{Andrea Celentano (PI).} He got is PhD in Physics in March 2014, with the thesis ``The Forward Tagger detector for CLAS12 at Jefferson Laboratory and the MesonEx experiment''. The main goal was to design the Forward Tagger Calorimeter detector for the CLAS12 experiment at Jefferson Laboratory \cite{FT}, in particular the front-end electronics and the readout system. He has an advanced expertise on organic scintillators, both solids and liquids, on photo-detectors (photomultiplies, SiPMs, APDs), and on the associated readout electronics, maturated during the PhD and the Master Thesis (measurement of the neutron yield from a thick Berillium target, with the time-of-flight technique, using plastic and liquid scintillator counters).

He is currently hired as a Post-Doc at INFN-Genova. He is involved the MesonEx experiment, and he is contributing to the detector development, following in particular the Data Acquisition System and the Slow Controls system. He is also participating to the Heavy Photon Search experiment (HPS) at Jefferson Laboratory \cite{hps}, where he is in charge of the design, construction, and operation of the Led Monitoring System for the electromagnetic calorimeter and of the related online monitoring software. He is co-spokeperson of the recently proposed Beam Dump eXperiment (BDX)
at Jefferson Laboratory \cite{BDX}. He is co-author of 37 papers published in peer-reviewed journals.
\\

\paragraph{Patrizia Boccacci.} She is Associate Professor at the Department of Computer Science of Genoa University (DIBRIS), associate to INFN-Genova Group V. 
Her main research interest is in numerical methods for the solution of inverse ill-posed problem and their applications, such as optical tomography for the investigation of crystal growth in microgravity, confocal microscopy, particle sizing, seismic tomography and astrophysics. More recently her main research interest is in image restoration problems related to the LBT (Large Binocular Telescope) project and in medical imaging, in particular SPECT (single photon emission computed tomography) and microwave tomography. She is co-author of several articles and co-author of one book. 


\subsubsection{DCCI-Unige unit}

The members of this unit are researchers at the Department of Chemistry of Genoa University, with a strong background on organic materials synthesis and characterization. The participation of the DCCI-Unige unit to this project is specific to the study of possible new organic scintillators with enhanced properties matched to this application, for example by doping existing liquid scintillators, or by introducing specific interfaces between the scintillator and the photo-detector to minimize diffusion effects. Therefore, the DCCI-Unige will be mainly involved during the last part of the project, i.e. in the Working Package 3.6. 
Nevertheless, thanks to the strong expertise of the members in organic materials, the participation of the DCCI-Unige unit will strengthen the whole \projname project. For example, the DCCI-Unige unit can provide a valuable support during the MonteCarlo simulations development, helping in the parametrization of the scintillator inhomogeneities and defects. 

\subsubsection{External support}

To increase  the success rate the project, we already contacted world-wide experts that will provide an external support on specific tasks, depending on the results that will be obtained during the first tests. We received a declaration of interest from Dr. Igor Nemchecko from the Dubna Joint Institute for Nuclear Research for support during the development of a specific organic scintillator, Dr. Marco Contalbrigo from INFN-Ferrara for the use of the MAROC3 ASIC, and Dr. Angelo Rivetti and Dr. Manuel Dionisio Da Rocha Rolo from INFN-Torino for the use of the TOFPET ASIC. 

\begin{table}[t]
\centering
\begin{tabular}{|c|c|c|c|}
\hline
\textbf{Name} & \textbf{Age} & \textbf{Position} & \textbf{Percentage of involvment} \\
\hline
\hline
\textbf{INFN-Ge unit} & & & \\
\hline
Andrea Celentano (PI) & 28 & Post-doc & 100$\%$ \\
\hline
Patrizia Boccacci & 59 & Associate Professor & $10\%$\\
\hline\hline
\textbf{DCCI-Unige unit} & & & \\
\hline
Davide Comoretto (Coordinator) & 51 & Researcher & 10$\%$ \\
\hline
Maila Castellano & 45 & Researcher & 10$\%$ \\
\hline
\end{tabular}
\caption{\small Composition of the two Research Units involved in the \projname project.}
\end{table}

\newpage

\section{Work plan (time-line and milestones)}

\begin{figure}[h]
\centering
\includegraphics[width=1.2\textwidth]{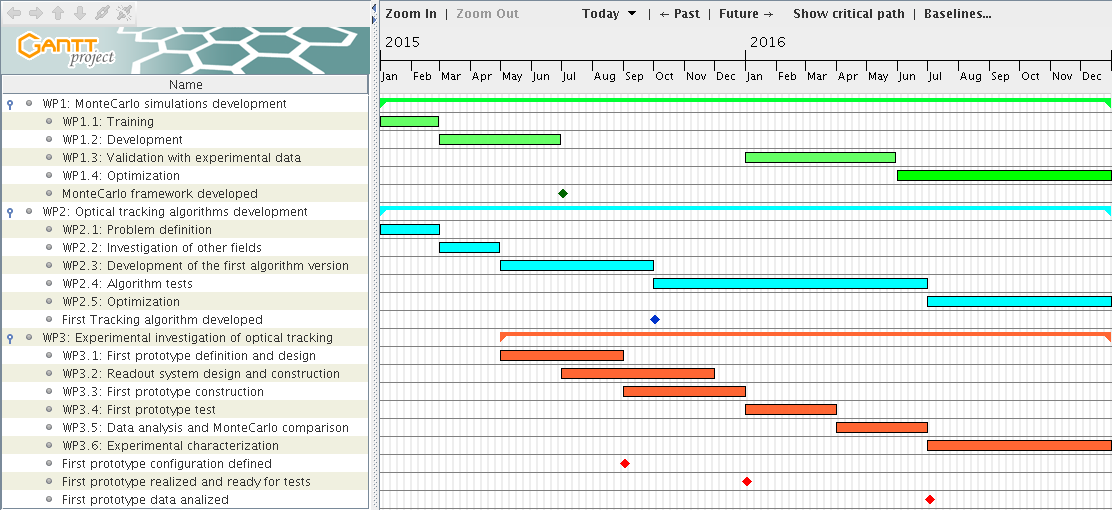}
\caption{\small \label{fig:gant} Gantt chart of the \projname project.}
\end{figure}

The detailed description of the project work plan, in terms of the different Working~Packages tasks, goals, milestones, and time-line  was reported in the previous section. The Gantt chart in figure \ref{fig:gant} summarizes it. For practical reasons, the project start date here reported is January 1$^{st}$, 2015.

\section{Impact}

The possibility to make an optical tracking detector using organic scintillators would be a major breakthrough in particle detectors, with a very large impact on many fields related to particle and nuclear physics. Here, we just mention a non-exhaustive list of the most promising applications:
\begin{description}
\item[On-beam experiments:]{ being the light the fastest way to transport information within a material, this application would permit to develop a detector intrinsically capable of sustaining a very high rate, limited only by the photo-detectors and the readout system.}
\item[Rare physics experiments:]{ these can benefit from the new optical tracking method, since it provides a convenient way to develop large-scale detectors with enhanced particle identification and background rejection capabilities, without any major technical issue related to the operations.}
\end{description}

As a concrete example, we briefly discuss the impact of this new technology on the aforementioned ``Beam Dump eXperiment'' (BDX) at Jefferson Laboratory, searching for light (MeV-GeV) Dark Matter \cite{BDX}.
The experiment is performed by placing a detector downstream of one of the JLab experimental Halls to measure DM particles that could be produced by the electron beam in the dump, pass through surrounding shielding material, and deposit visible energy inside the detector by scattering off detector nuclei or (if unstable) by decaying inside the detector volume. The highest energy transfer, and thus the highest signal in the detector, is obtained with low Z nuclei, and is maximum for the scattering on protons. For this reason, a plastic-scintillator based detector represents a very convenient solution, due to the high H/C ratio, the high density, required to reach a sufficient luminosity, and the high light yield, necessary to detect low-energy scattered particles.

The main limiting factor to the experimental sensitivity are the cosmogenic backgrounds, i.e. cosmic muons and neutrons, not properly shielded or vetoed, hitting the detector and releasing a visible amount of energy. In particular, non-trivial background sources are represented by cosmic muons being captured or decaying in the detector, or in the surrounding material. To maximize the experimental reach, these must be clearly identified and rejected.

This technology would add to the detector particle tracking and identification capability, enhancing the experimental reach. Preliminary simulations showed that, in a very conservative approach, an optical tracker with 1 mm spatial resolution could identify $\simeq 95\%$ of the muons-induced signals, and reduce the neutrons-induced background by $\simeq 33\%$ . This would result in a background rejection factor of $\simeq  10$. 

Furthermore, the measurement of the kinematics of the recoiling particle opens new possibilities and, in case of a positive signal, permits to further explore dark matter properties. 

\begin{itemize}
\item{In case of metastable dark matter particles are produced in the beam and decay to visible matter inside the detector, the mass of these can be reconstructed by measuring the daughters kinematics.}
\item{For a narrow range of parameters, this technology could also give a handle on the lifetime of a dark-matter excited state, if the de-excitation is displaced and has some probability of decaying inside the detector, but away from the interaction point.}

\end{itemize}

\section{Outcome and outlook}


The development of an optical tracking detector using organic scintillators is a very challenging task. \textbf{Our ultimate goal is to test this new technology, and to develop, as proof-of-principle, a working prototype.} Nevertheless, it is worth stressing that, independently from the final result that will be obtained at the end of the project, each of the Working Packages will provide an outcome that represents, by itself, a new significant result:
\begin{itemize}
\item{We will optimize the description of optical processes in Geant4, on the basis of the measurements that we will perform with the prototype.}
\item{For each component of the optical tracker detector, we will identify the critical properties and their influence on the performance.}
\end{itemize}

The latter point is the pre-requisite for the natural evolution of this project towards a mature technology that can be employed in an experiment: the design and construction an optical tracking detector making use of specifically designed components. \textbf{The results of this project will determine which properties these components should have, given the requirements of a specific application.}
Here, we just outline what is the extensive R\&D activity that must be performed to reach this ambitious goal.

\begin{itemize}
\item{The design performances of LA-PPD photo-detectors, currently under development, are perfectly matched to the requirements of this project. Whenever the first samples will be available, they should be tested within this context, and possibly be adopted as a new, innovative photo-detector solution.}
\item{The design of a custom ASIC will permit to realize a readout electronics system exactly matched to the detector requirements.}
\item{The tracking reconstruction algorithm is a critical component of the project. Here, we focused on the definition of the problem: our goal is to develop a first, preliminary version of the tracking reconstruction. The next step would be to optimize it, focusing on the involved approximations, the computational resources, and the execution time. The ultimate goal is to run the algorithm \textit{on-line}, on a computational system (FPGA/GPU) connected to the readout system.}
\end{itemize}

\addcontentsline{toc}{chapter}{Bibliography}
\printbibliography

\end{document}